\documentclass[a4paper,12pt]{article}
\usepackage{amsmath}
\usepackage{amssymb}
\usepackage{listings}
\usepackage{comment}
\usepackage[pdftex]{graphicx}
\usepackage{subfigure}
\begin{document}

\title{The maximum force in a column under constant speed compression}

\author{Vitaly A. Kuzkin\footnote{Vitaly A. Kuzkin,
              Peter the Great Saint Petersburg Polytechnical University,
              Polytechnicheskaya st. 29, Saint Petersburg, Russia;
              Institute for Problems in Mechanical Engineering RAS, Bolshoy pr. V.O. 61, Saint Petersburg, Russia
              Tel.: +7-981-7078702; e-mail: kuzkinva@gmail.com;
           Mona M. Dannert,              Institute of Mechanics and Computational Mechanics,
Leibniz University Hanover,
Appelstrasse 9a, 30167 Hannover
}        \and
        Mona M. Dannert
}

\maketitle

\begin{abstract}
Dynamic buckling of an elastic column under compression at constant speed is investigated assuming the first-mode buckling. Two cases are considered: (i) an imperfect column~(Hoff's statement), and (ii) a perfect column having an initial lateral deflection. The range of parameters, where the maximum load supported by a column exceeds Euler static force is determined. In this range, the maximum load is represented as a function of the compression rate, slenderness ratio, and imperfection/initial deflection. Considering the results we answer the following question: ``How slowly the column should be compressed in order to measure static load-bearing capacity?'' This question is important for the proper setup of laboratory experiments and computer simulations of buckling.

Additionally, it is shown that the behavior of a perfect column having an initial deflection differ significantlys form the behavior of an imperfect column. In particular, the dependence of the maximum force on the compression rate is non-monotonic. The analytical results are supported by numerical simulations and available experimental data.

\end{abstract}

\section{Introduction}

Buckling of columns~(rods, beams) under compression is a classical problem for mechanics of solids. In 1744 Leonard Euler predicted the critical buckling force for a compressed column in statics. Numerous experimental and theoretical studies have revealed that the behavior of a column in dynamics is significantly more complicated. In particular, in dynamics the maximum force is usually not equal to Euler static force~\cite{Tovstik 2013,Hoff_51}. Moreover dynamic buckling behavior significantly depends on the way of compression. A review of different loading conditions is given, for example, in paper~\cite{Karagiozova}. Buckling under the action of time-dependent aperiodic load is considered in papers~\cite{Kornev19741,Kornev19742,Markin}. Sudden application of a constant force, refereed to as Ishlinsky-Lavrentiev problem~\cite{Lavrentev}, was investigated, for example, in papers~\cite{Tovstik 2013,Belyaev Dokl 2013,Belyaev MTT 2013,Morozov_Tovstik 2014,BelMorTovTov_2015}. Buckling under impact loading was studied theoretically and experimentally in papers~\cite{Waas,Mimura 2008}.

In the above-mentioned loading regimes the transition to quasi-statics is not straightforward or even impossible. Therefore static load-bearing capacity is measured using hydraulic testing machines.   In this case column ends move towards each other with constant velocity. The longitudinal force in the column increases with time. At some moment of time the force reaches the maximum value, further refereed to as ``the maximum force supported by a column''. If the velocity is sufficiently small, then the quasi-static behavior is expected. However in 1951 N.J. Hoff has shown that even at very small~(compared to sound speed) compression rates, the maximum force supported by a column significantly exceed Euler static force~\cite{Hoff_51}. This theoretical result is supported by  experimental observations~\cite{Hoff_exp,Mimura 2012}. In particular, in paper~\cite{Mimura 2012} it has been shown that for thin columns the maximum force can exceed Euler static force by a factor of~$100$. The difference is caused by lateral inertia of a column~\cite{Hoff_51}. Assuming the first mode buckling, Hoff has demonstrated that the maximum force is a function of two parameters: (i) an amplitude of column's imperfection, and (ii) a similarity number, depending on length/thickness ratio of the column and the compression rate.
Buckling under constant speed compression was studied in many papers~\cite{Motamarri,Sevin 1960,Dym_68,Elishakoff,Kounadis,Tyler}. The influence of axial inertia~\cite{Sevin 1960}, random imperfection~\cite{Elishakoff}, and boundary conditions~\cite{Motamarri} was investigated. In particular, it has been shown that in the case of small imperfections the time evolution of column's  deflection is represented in terms of Bessel functions~\cite{Hoff_51}, Lomel functions~\cite{Elishakoff} or, Airy functions~\cite{Tyler}.  However, to our knowledge, analytical expression for the maximum load supported by a column as a function of compression rate, and imperfection of the column has not been reported.

In the present paper we consider dynamic buckling of a column under constant speed compression.  The paper is organized as follows. In section~\ref{chapt: Hoff}, we recall Hoff's formulation of buckling problem for an imperfect column~\cite{Hoff_51}. An example, demonstrating typical behavior of the system is given. In section~\ref{chapt: Analyt solution}, slightly imperfect and perfect columns are considered. In both cases, simple analytical expressions for the maximum  force as a function of compression rate, slenderness ration, and initial deflection/imperfection are derived. Using the expressions, we answer the following question: ``How slowly the column should be compressed in order to measure static load-bearing capacity?'' Analytical results are compared with corresponding numerical solutions and experimental data~\cite{Hoff_exp}. In conclusions, the importance of strain-rate effects in buckling of micro and nano columns is discussed.

\section{Dynamic buckling of an imperfect column: Hoff's statement}\label{chapt: Hoff}

\begin{figure*}[htb]%
\begin{center}
\includegraphics*[scale=0.5]{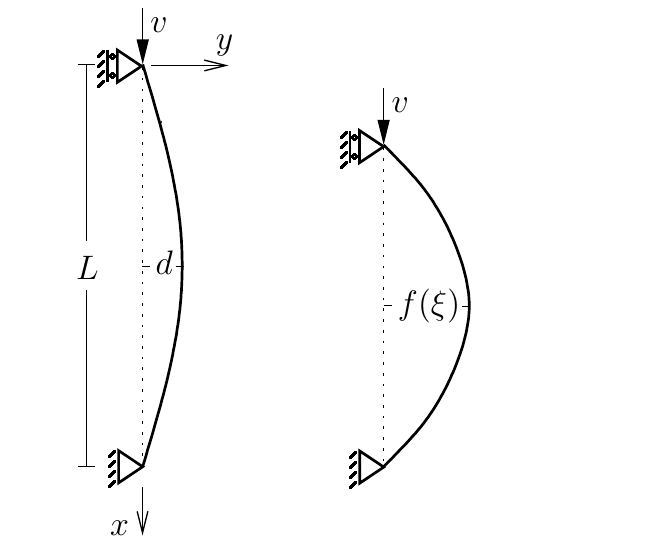}
\caption{Initial~(left) and current~(right) configurations of the column. Here~$d$ is the amplitude of imperfection.}
\label{modelHoff}
\end{center}
\end{figure*}

In the present section, we recall the formulation of Hoff's problem~\cite{Hoff_51,Dym_68,Elishakoff}.
Hoff has considered compression of naturally curved column in a hydraulic testing machine. Column ends
move towards each other with constant velocity~$v$~(see figure~\ref{modelHoff}).
Longitudinal vibrations of the column are neglected. In paper~\cite{Sevin 1960} it is shown that at sufficiently small compression rates the influence of longitudinal vibrations is insignificant.
The lateral deflection of the column~$y(x,t)$ satisfies the following partial differential equation~\cite{Dym_68}:
\begin{equation}\label{eq: general transverse motion}
 EI \frac{\partial^4 (y-y_0)}{\partial x^4}  + P \frac{\partial^2 y}{\partial x^2} + \rho A \frac{\partial^2 y}{\partial t^2} = 0,
\end{equation}
where~$y_0(x)$ is the lateral deflection due to the imperfection, $P$ is the longitudinal force, $E$ is Young's modulus, $I$ is the moment of inertia of the cross-section, $A$ is the cross-sectional area,
and $\rho$ is the density. It is assumed that the longitudinal force in constant along the column.
Summing the contributions of longitudinal displacement, $vt$, and lateral deflection,
yields the following expression for the longitudinal force:
\begin{equation}\label{eq: general compression force}
P = \frac{EA}{L} \left(vt - \frac{1}{2} \int\limits_0^L \left[ \left(\frac{\partial y}{\partial x}\right)^2 - \left(\frac{\partial y_0}{\partial x}\right)^2 \right] {\rm d} \! x \right).
\end{equation}
where $L$ is the column's length.
The following boundary conditions are used
\begin{equation}
\left. y \right|_{x = 0} = \left. y \right|_{x = L} = 0, \quad \quad \left. \frac{\partial^2 y}{\partial x^2} \right|_{x=0} = \left. \frac{\partial^2 y}{\partial x^2} \right|_{x=L} = 0.
\end{equation}
Hoff has assumed the first-mode buckling. This assumption is satisfied at sufficiently low compression speeds.
Then the deflection can be approximated as
\begin{equation}\label{eq.: deflection}
y(x,t) = R f(\xi) \sin \frac{\pi x}{L}, \quad \quad \xi = \frac{v t L}{\pi^2 R^2},
\end{equation}
where $\xi$ is a dimensionless time, $R = \sqrt{I/A}$ is the radius of gyration. In Hoff's statement the initial deflection is caused by imperfection of the column. The imperfection is also approximated by a sine wave:
\begin{equation}
y(x,0)=y_0(x)=R d \sin \frac{\pi x}{L},
\end{equation}
where~$d$ is a dimensionless amplitude of the imperfection.
Inserting~(\ref{eq: general compression force}) and~(\ref{eq.: deflection}) into equation~(\ref{eq: general transverse motion}) yields an
 ordinary differential equation for dimensionless deflection amplitude~$f(\xi)$:
\begin{equation}\label{diff eq f}
f'' + \Omega \left[\left(1-\xi- \frac{d^2}{4}\right) f + \frac{1}{4}f^3    - d \right] = 0, \quad \quad
f(0) = d, \quad \quad f'(0) = 0.
\end{equation}
where prime denotes the derivative with respect to dimensionless time~$\xi$.
Equation~(\ref{diff eq f}) shows that  buckling behavior is governed by two parameters: the dimensionless imperfection amplitude,~$d$, and Hoff's similarity number,~$\Omega$, defined as
\begin{equation}
\Omega = \pi^8 \left( \frac{R}{L} \right)^6 \left( \frac{v_s}{v} \right)^2,
\end{equation}
where $v_s=\sqrt{E/\rho}$ is the velocity of longitudinal waves in a column. Note that $\Omega \rightarrow 0$ corresponds to rapid loading, while $\Omega\rightarrow +\infty$ corresponds to quasi-statics. Typical values of~$\Omega$ are given in section~\ref{chap.: imperf}.

The relationship between the longitudinal force, $P$, dimensionless time, $\xi$, and deflection,~$f$, follows
 from~(\ref{eq: general compression force}) and~(\ref{eq.: deflection}):
\begin{equation}\label{eq.: P by PE1}
\frac{P}{P_E} = \xi - \frac{1}{4}\left(f(\xi)^2 - d^2\right),
\end{equation}
where~$P_E = (\pi^2 EI)/L^2$ is Euler static force for a perfect column.

\begin{figure*}[htb]%
\begin{center}
\includegraphics[width=0.5\textwidth]{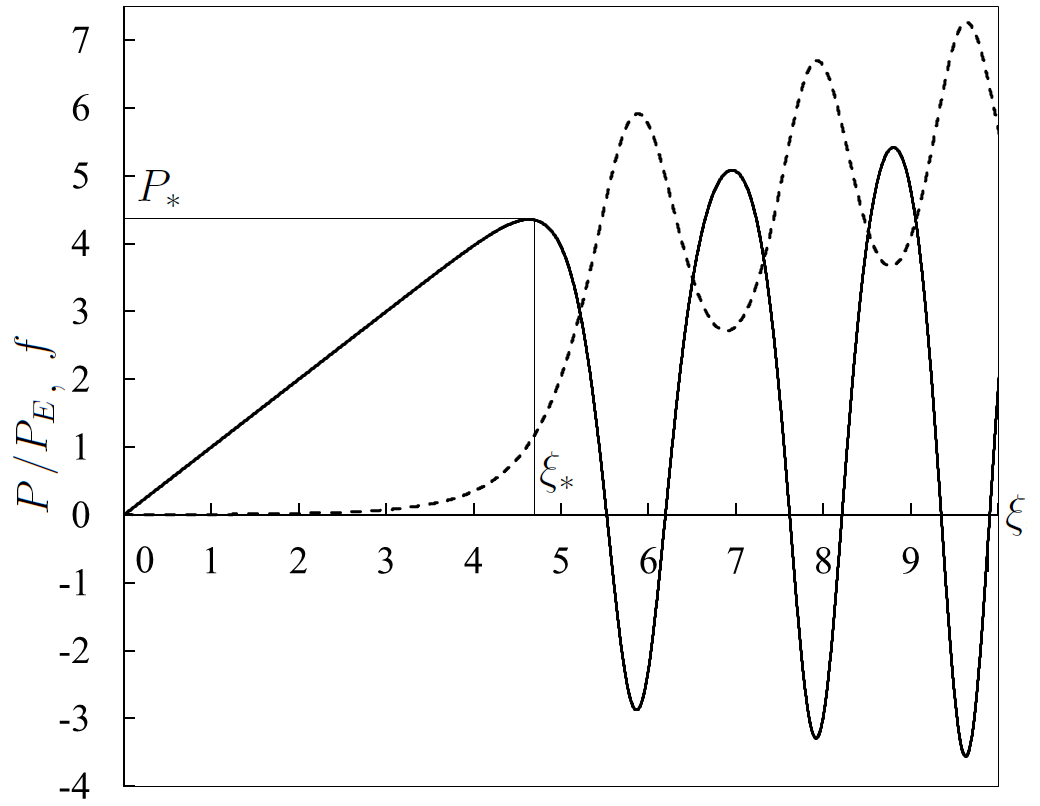}
\caption{Longitudinal force~(solid line), and deflection~(dashedline) of a column: numerical solution of equation~(\ref{diff eq f}) for~$\Omega=1, d=0.01$.}
\label{fig.: force}
\end{center}
\end{figure*}

As an example, dependencies of a longitudinal force and deflection on dimensionless time~$\xi$, obtained by numerical solution of equation~(\ref{diff eq f})
for~$\Omega=1, d=10^{-2}$, are shown in figure~\ref{fig.: force}.
It is seen that the force increases almost linearly with~$\xi$. It reaches Euler static force at~$\xi\approx 1-d^2/4$. For~$\xi>1-d^2/4$ the system has negative stiffness and the deflection rapidly increases. However because of inertia, a certain time is required for a column to buckle~\cite{Hoff_51}. We denote the moment of time when the longitudinal force takes the maximum value as~$\xi_*$. Then~$\xi_*$ is defined
by the following equation:
\begin{equation}\label{eq.: dPdxi}
\left. \frac{{\rm d}\!P}{{\rm d}\!\xi} \right|_{\xi = \xi_*} = 0 \quad \Rightarrow \quad f(\xi_*) f'(\xi_*)= 2.
\end{equation}
If the deflection~$f(\xi)$ is known, then solving equation~(\ref{eq.: dPdxi}) with respect to~$\xi_*$ and substituting the result into formula~(\ref{eq.: P by PE1}), yields
 the maximum force supported by a column~$P_*$:
\begin{equation}\label{eq.: P by PE}
\frac{P_*}{P_E} = \xi_* - \frac{1}{4}\left(f(\xi_*)^2 - d^2\right),
\end{equation}
Thus, the dependence of maximum force on similarity number,~$\Omega$, and imperfection,~$d$, is given in the implicit form by equations~(\ref{diff eq f}), (\ref{eq.: dPdxi}), and~(\ref{eq.: P by PE}). The explicit dependence is derived in the next section.

\section{Calculation of the maximum force supported by a column}\label{chapt: Analyt solution}
\subsection{Time required for buckling}
Equation~(\ref{eq.: dPdxi}) shows that because of lateral inertia, a certain time is required for a column to buckle. This was first observed by Hoff~\cite{Hoff_51}: ``{\it During a rapid loading the transverse motion of the elements of the column is retarded by inertia of their masses. For this reason the dynamic deflection lag behind the values that correspond to infinitely slow loading}.'' In the present section, we derive the expression for the time,~$\xi_*$, corresponding to the maximum value of the force, as a function of Hoff's similarity number, $\Omega$, initial deflection, $f_0$, and imperfection amplitude,~$d$.

Consider the case of small deflections~($f \ll 1$).
Then neglecting the cubic term in Hoff's differential equation~(\ref{diff eq f}) yields:
\begin{equation}\label{eq.: Hoff neglect}
f'' + \Omega \left[\left(1-\xi- \frac{d^2}{4}\right) f - d \right] = 0, \quad \quad
f(0) = f_0, \quad \quad f'(0) = 0.
\end{equation}
Note that, in general, we assume~$f_0 \neq d$. This allows us to consider a perfectly shaped column~($d=0$). The solution of equation~(\ref{eq.: Hoff neglect}) is represented in terms of Airy's functions:
\begin{equation}\label{solve lin diff}
\begin{array}{l}
f(\zeta) = \pi f_0 \left[ {\rm Ai}(\zeta){\rm Bi}' \left(-\Omega^{\frac{1}{3}} \right) - {\rm Ai}' \left(-\Omega^{\frac{1}{3}} \right) {\rm Bi}(\zeta)
 \right] \\[4mm]
  \quad \quad - \pi d \, \Omega^{\frac{1}{3}} \left[{\rm Ai}(\zeta) \int\limits_{-\Omega^\frac{1}{3}}^{\zeta} {\rm Bi}(z) {\rm d}\!z - {\rm Bi}(\zeta) \int\limits_{-\Omega^\frac{1}{3}}^{\zeta} {\rm Ai}(z) {\rm d}\!z \right], \qquad \zeta =\Omega^{\frac{1}{3}} \left(\xi + \frac{d^2}{4}-1\right),
\end{array}
\end{equation}
where the identity~${\rm Bi}'{\rm Ai}-{\rm Ai}'{\rm Bi}=1/\pi$ was used. Properties of Airy's functions are described, for example, in a book~\cite{Olver_10}.

From the maximum condition~(\ref{eq.: dPdxi}) it follows that for small initial deflection,~$f_0$, and imperfection,~$d$,
the parameter~$\zeta_*=\Omega^{\frac{1}{3}} (\xi_* + d^2/4 - 1)$ is large. Then neglecting in formula~(\ref{solve lin diff}) the contribution of function~${\rm Ai}$ and substituting the resulting expression into~(\ref{eq.: dPdxi}), yields
 the equation relating~$\xi_*$ and~$\Omega$:
\begin{equation}\label{eq: xi imp}
 \pi^2 \, \Omega^{\frac{1}{3}} \, \left[ f_0 \, {\rm Ai}'\left(-\Omega^{\frac{1}{3}}\right)
 - d \, \Omega^{\frac{1}{3}} \int\limits_{-\Omega^{\frac{1}{3}}}^{\zeta_*} {\rm} {\rm Ai}(z) {\rm d}z
  \right]^2 {\rm Bi}(\zeta_*){\rm Bi}'(\zeta_*) = 2, \qquad \zeta_* =\Omega^{\frac{1}{3}} \left(\xi_*+ \frac{d^2}{4}-1\right).
\end{equation}
The key step is to use the following
asymptotic formulas for Airy functions at~$\zeta_* \rightarrow +\infty$:
\begin{equation}\label{eq: BiBi'}
\int\limits_{-\Omega^{\frac{1}{3}}}^{\zeta_*} {\rm} {\rm Ai}(z) {\rm d}z \sim
 \int\limits_{-\Omega^{\frac{1}{3}}}^{0} {\rm} {\rm Ai}(s) {\rm d}s + \frac{1}{3},
 \qquad {\rm Bi}(\zeta_*) \sim \frac{ e^{\frac{2}{3} \zeta_*^{\frac{3}{2}}} }{ \sqrt{\pi} \zeta_*^{\frac{1}{4}} },
\qquad
{\rm Bi}'(\zeta_*) \sim \frac{\zeta_*^{\frac{1}{4}}}{\sqrt{\pi}}e^{\frac{2}{3} \zeta_*^{\frac{3}{2}}}
\end{equation}
%
Substituting the asymptotic formulas~(\ref{eq: BiBi'}) into~(\ref{eq: xi imp}) and solving the resulting equation with
respect to~$\xi_*$ yields:
\begin{equation}\label{xi_star}
\xi_* = 1 -\frac{d^2}{4} +
	\left\lbrace
		\frac{3}{4 \, \Omega^{\frac{1}{2}}} \ln
			\left[
				\frac{2}{\pi\Omega^{\frac{1}{3}}}
					\left(
						f_0 {\rm Ai}'\left(-\Omega^{\frac{1}{3}}\right)
 - d \, \Omega^{\frac{1}{3}} \left(\int\limits_{-\Omega^{\frac{1}{3}}}^{0} {\rm Ai}(z) {\rm d}z + \frac{1}{3}\right)
 					\right)^{-2}
			\right]
		\right\rbrace^{\frac{2}{3}}.
\end{equation}
Formula~(\ref{xi_star}) gives the explicit dependence of the ``time required for buckling''  on the similarity number, imperfection amplitude, and initial deflection. It is applicable to perfect and imperfect columns. Further these two cases are considered separately.

\subsection{Imperfect column}\label{chap.: imperf}
Following Hoff, we assume that initial deflection of the column is due to imperfection only~($f(0)=d$). From asymptotic formulas~(\ref{eq: BiBi'}) it follows that for small imperfections,~$d$,  the term~$f(\xi_*)^2/4$ in equation~(\ref{eq.: P by PE}) can be  neglected and therefore~$P_*/P_E \approx \xi_* +d^2/4$. Also in order to simplify formula~(\ref{xi_star}), we assume that~$\Omega$ is large. In this case~${\rm Ai}'\left(-\Omega^{\frac{1}{3}}\right)$ can be neglected compared to~$\Omega^{\frac{1}{3}}$ and
$\int_{-\Omega^{\frac{1}{3}}}^{0} {\rm Ai}(s) {\rm d}s \sim 2/3$.
Then the expression for the maximum force, $P_*$, takes the form:
\begin{equation}\label{eq.: solve imperfect}
\frac{P_*}{P_E} = 1 + \left( \frac{3}{4 \, \Omega^\frac{1}{2}} \ln \left[\frac{2}{\pi \, d^2 \, \Omega} \right] \right)^{\frac{2}{3}}, \quad \quad \Omega = \pi^8 \left( \frac{R}{L} \right)^6 \left( \frac{v_s}{v} \right)^2.
\end{equation}
Formula~(\ref{eq.: solve imperfect}) gives simple analytical relationship between the main parameters of the problem: maximum force, compression rate, length/thickness ratio, and imperfection of the column. This is the main result of the present paper.

\begin{figure*}[htb]%
\begin{center}
\includegraphics[width=0.5\textwidth]{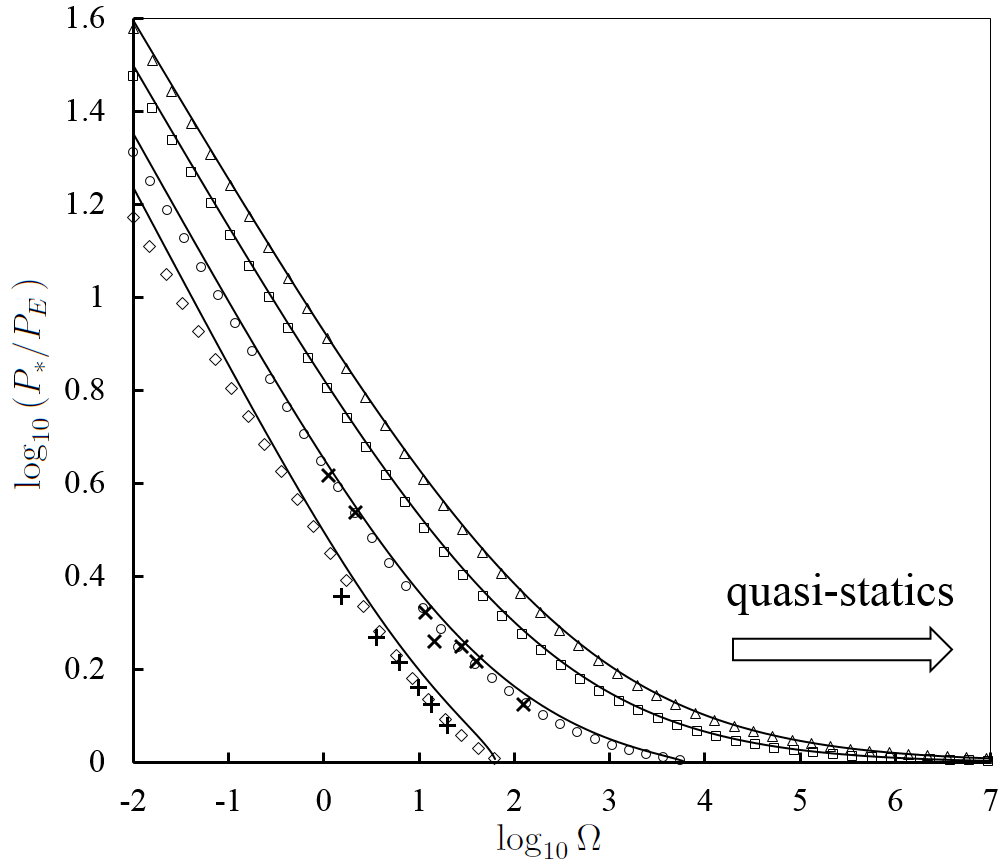}
\caption{Comparison of analytical~(solid lines), numerical, and experimental results for different imperfections: $d = 10^{-1}$~(diamonds --- numerical; pluses --- experimental~\cite{Hoff_exp}), $d = 10^{-2}$~(circles --- numerical, crosses --- experimental~\cite{Hoff_exp}),  $d = 10^{-4}$ (squares --- numerical), and  $d = 10^{-6}$ (triangles --- numerical).}
\label{fig.: dF_imperf}
\end{center}
\end{figure*}

Formula~(\ref{eq.: solve imperfect}) shows that the influence of compression rate on the maximum force is very strong.
The influence of imperfection is logarithmic~(see figure~\ref{fig.: dF_imperf}).  The maximum force is equal
to Euler static force for a certain value of similarity number,~$\Omega_E$,
depending on the imperfection:
\begin{equation}\label{Omega_E}
  P_*=P_E \quad \Rightarrow \quad  \Omega =\Omega_E = \frac{2}{\pi d^2}.
\end{equation}
Formula~(\ref{eq.: solve imperfect}) is applicable for~$\Omega \leq \Omega_E$. In this interval, the maximum force is larger than Euler static force. Formula~(\ref{Omega_E}) can be used, for example, for choosing appropriate compression rate in laboratory experiments and numerical simulations. According to formula~(\ref{Omega_E}),
in order to measure the static load-bearing capacity, the velocity of compression should satisfy the following inequality:
\begin{equation}\label{ineq}
 \frac{v}{v_s} < \frac{\pi^{\frac{9}{2}} d}{\sqrt{2}} \left(\frac{R}{L}\right)^3.
\end{equation}

To judge the accuracy of formula~(\ref{eq.: solve imperfect}), we compare the results with numerical solution of Hoff's non-linear
equation~(\ref{diff eq f}). Leap-frog integration scheme~\cite{Verlet} is used. The dimensionless time steps for the numerical integration is~$\Delta \xi=0.016 \Omega^{-1/2}$. Note that the time step satisfies the stability condition for numerical scheme~$\Delta \xi < \frac{\pi}{10} \, \Omega^{-1/2}$.  In the following simulations Hoff's similarity number belongs to the interval~$\Omega\in[10^{-2}; 10^{7}]$. In order to give more intuitive numbers, consider the column with length/thickness ratio~$L/R=10^2$. Then compression velocities are in the range~$v/v_s \in [3 \cdot 10^{-8}; 10^{-3}]$. So for a steel column with~$v_s \approx 5 \cdot 10^3$ m/s, the compression velocities from~$0.15$~mm/s to $5$~m/s are considered.

Compare the maximum force supported by a column,~$P_*$, calculated using expression~(\ref{eq.: solve imperfect})
with the results of numerical solution of equation~(\ref{diff eq f}). The results for imperfections~$d = 10^{-1}, 10^{-2}, 10^{-4}, 10^{-6}$ are
shown in figure~\ref{fig.: dF_imperf}. The maximum value of~$\Omega$ considered below is equal to~$\Omega_E$. Experimental
points from the papers~\cite{Hoff_exp,Hoff_65} are added~($d = 10^{-1}$ (stars) and $d = 10^{-2}$ (pluses)).
It is seen that predictions of analytical formula~(\ref{eq.: solve imperfect}) are in a good agreement with experimental and numerical results. As expected, the accuracy of formula~(\ref{eq.: solve imperfect}) increases with decreasing imperfection~($d\rightarrow 0$). In spite of the fact that formula~(\ref{eq.: solve imperfect}) is derived assuming large~$\Omega$, it has acceptable accuracy in the whole range of similarity numbers considered above.

\subsection{Perfect column}\label{chap.: perf}

Calculate the maximum force supported by a perfect column~($d=0$).
Substituting formula~(\ref{xi_star}) into
formula~(\ref{eq.: P by PE}) and neglecting the term~$f(\xi_*)^2$, yields
\begin{equation}\label{eq.: solve perfect}
\frac{P_*}{P_E} = 1 + \left( \frac{3}{4 \Omega^{\frac{1}{2}}} \ln \left[\frac{2}{\pi f_0^2 \, \Omega^{\frac{1}{3}} {\rm Ai}'\left(-\Omega^{\frac{1}{3}}\right)^2 } \right] \right)^{\frac{2}{3}},
\end{equation}
Formula~(\ref{eq.: solve perfect}) gives an explicit dependence of the maximum force supported by a perfect column on similarity number and initial deflection. Consider asymptotic expression for~${\rm Ai'\left(-\Omega^{\frac{1}{3}}\right)}$ at large values of~$\Omega$:
\begin{equation}\label{Ai_prime}
 {\rm Ai}'\left(-\Omega^{\frac{1}{3}}\right) \sim - \frac{\Omega^{\frac{1}{12}}}{\sqrt{\pi}} \cos\left(\frac{2}{3}\Omega^{\frac{1}{2}} + \frac{\pi}{4}\right)
\end{equation}
From formulas~(\ref{eq.: solve perfect}), (\ref{Ai_prime}) it follows that the the dependence of the maximum force on~$\Omega$ for a perfect column is non-monotonic. Moreover, the maximum force tends to infinity at similarity numbers, $\Omega_k$, satisfying the equation~${\rm Ai}'\left(-\Omega_k^{\frac{1}{3}}\right)=0$. Using asymptotic formula~(\ref{Ai_prime}),
yields an approximate expression for~$\Omega_k$:
\begin{equation}\label{eq.: omega_k}
 \Omega_k \approx \frac{9\pi^2}{64
 }\left(2k-1\right)^2, \qquad k=1,2,...
\end{equation}
The dependence of column's deflection on time for~$\Omega=\Omega_k$ follows from~(\ref{solve lin diff}):
\begin{equation}\label{f om_k}
\begin{array}{l}
f(\xi) = \pi f_0 {\rm Bi}' \left(-\Omega_k^{\frac{1}{3}} \right){\rm Ai}\left(\Omega_k^{\frac{1}{3}} (\xi-1)\right).
\end{array}
\end{equation}
Formula~(\ref{f om_k}) shows that the deflection tends to zero as time tends to infinity. Therefore the solution of linear equation~(\ref{eq.: Hoff neglect}) predicts that for similarity numbers~$\Omega_k$ there is no buckling and the longitudinal force increases infinitely.

\begin{figure*}[htb]%
\begin{center}
\includegraphics[width=0.5\textwidth]{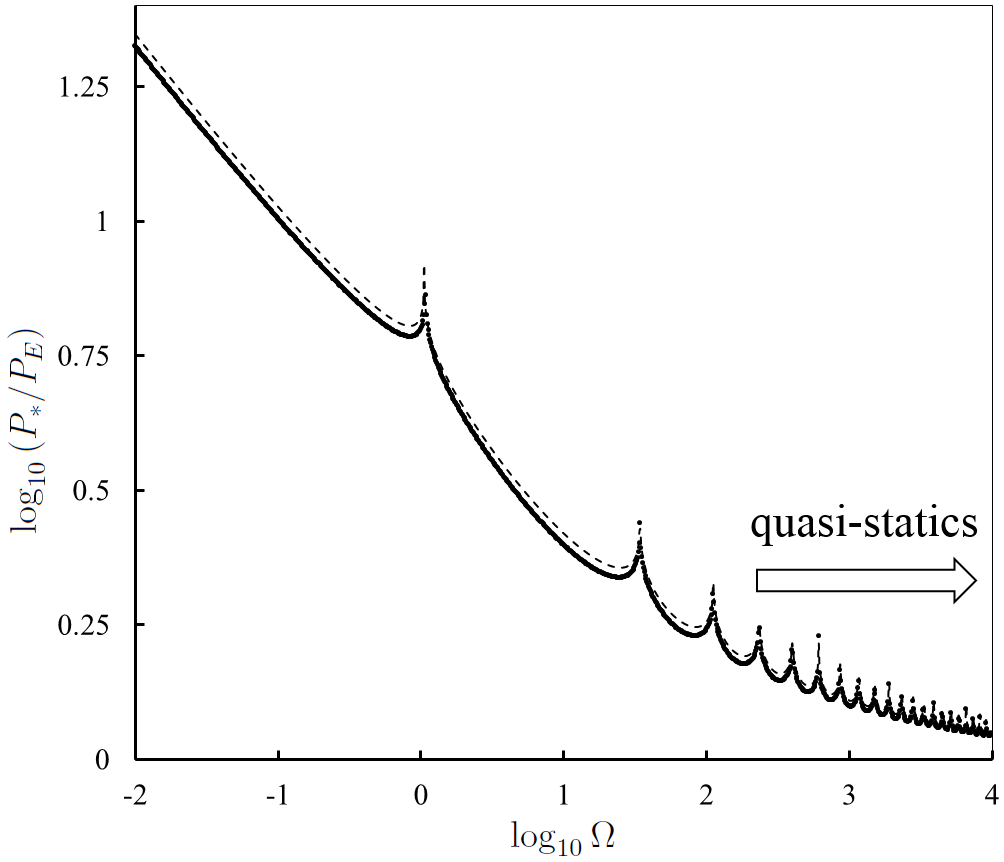}
\caption{Dependence of the maximum force on the similarity number for a perfect column: analytical~(dashed line) and numerical~(points) results for $f_0 = 10^{-3}$. The line corresponding to numerical solution contains~$10^3$ points.}
\label{fig.: perfSingle}
\end{center}
\end{figure*}

Compare the dependence of the maximum force on similarity number~(\ref{eq.: solve perfect}) with results of numerical solution of nonlinear Hoff's equation.
The results for~$f_0=10^{-3}$ are  shown in figure~\ref{fig.: perfSingle}.
It is seen that both analytical and numerical solutions have maxima at  similarity numbers given by approximate formula~(\ref{eq.: omega_k}). Therefore the behavior of a perfect column having initial deflection is significantly different than the behavior of an imperfect column.

\section{Conclusion}

Dynamic buckling of an elastic column under constant speed compression was considered assuming the first mode buckling.
Perfect~(straight) and imperfect~(naturally curved) columns were investigated. For imperfect columns the range of parameters, where the buckling force exceeds Euler static force was determined analytically. In this range,  the dependence of maximum force supported by a column on compression rate, length/thickness ratio, and imperfection is described by simple analytical formula~(\ref{eq.: solve imperfect}). Formula~(\ref{eq.: solve imperfect}) shows that the dependence of the maximum force on compression rate and length/thickness ratio is very strong. At the same time, the influence of imperfection is logarithmic. Numerical simulations show that formula~(\ref{eq.: solve imperfect}) has acceptable accuracy in the range of parameters, where the maximum force exceeds Euler force. The accuracy increases with decreasing amplitude of column's imperfection. These theoretical results suggest that the compression rate, required for measuring static load-bearing capacity of columns should satisfy the inequality~(\ref{ineq}). This inequality can be used for proper setup of laboratory experiments and computer simulations of buckling. For a perfect column the dependence of the maximum force on compression rate is non-monotonic. It has an infinite number of maxima at the points determined by formula~(\ref{eq.: omega_k}). The validity of analytical results is supported by corresponding numerical solutions.

Finally, let us note that investigation of perfect and nearly perfect structures is motivated by recent development of micro and nanotechnologies.
The results obtained in the present paper  provide some insights into buckling behavior of micro and nano structures, such as, nanowires~\cite{nanowires power law}, whiskers~\cite{whiskers}, nanotubes~\cite{nanutbes 2012,Korobeynikov}, etc. At nanoscale the column may have perfectly straight shape. Deviations from the straight configuration are due to thermal motion only. The results of the present paper suggests, that the influence of strain rate is very important and should be taken into account in computer simulations and real experiments on buckling of nanostructures.

The authors are deeply grateful to A.K. Belyaev, D.A. Indeytsev, E.A. Ivanova, A.M. Krivtsov, A.M. Linkov, A. Pshenov, and S. Rudykh for useful discussions. This work was financially supported by Russian Science Foundation.



\end{document}